\documentclass[12pt]{article}
\usepackage{latexsym}
\usepackage{epsfig}
\nonstopmode

\newtheorem{theorem}{Theorem}
\newtheorem{prop}[theorem]{Proposition}
\newtheorem{corollary}[theorem]{Corollary}

\newtheorem{lemma}[theorem]{Lemma}

\newcommand{\be}{\begin{equation}}
\newcommand{\ee}{\end{equation}}
\newcommand{\bea}{\begin{eqnarray}}
\newcommand{\eea}{\end{eqnarray}}
\newcommand{\beaa}{\begin{eqnarray*}}
\newcommand{\eeaa}{\end{eqnarray*}}
\newcommand{\barr}{\begin{array}}
\newcommand{\earr}{\end{array}}

\newcommand{\benum}{\begin{enumerate}}
\newcommand{\eenum}{\end{enumerate}}

\newcommand{\blem}{\begin{lemma}}
\newcommand{\elem}{\end{lemma}}
\newcommand{\bthm}{\begin{theorem}}
\newcommand{\ethm}{\end{theorem}}
\newcommand{\bcor}{\begin{corollary}}
\newcommand{\ecor}{\end{corollary}}
\newcommand{\bpr}{\begin{prop}}
\newcommand{\epr}{\end{prop}}
\newcommand{\nn}{\nonumber}
\newcommand{\nind}{\noindent}
\newcommand{\Proof}{\nind{\bf Proof:}\ \ }

\newcommand{\Reals}{{\mbox{\bf R}}}
\newcommand{\QED}{$\hfill\Box$}
\newcommand{\pa}{\partial}

\newcommand{\var}{\mbox{var}}

\newcommand{\De}{\Delta}
\newcommand{\de}{\delta}

\newcommand{\cE}{{\cal E}}

\newcommand{\cF}{{\cal F}}

\newcommand{\cG}{{\cal G}}

\newcommand{\bR}{{\bf R}}

\newcommand{\bQ}{{\bf Q}}

\begin{document}
\bibliographystyle{alpha}

\title{Pricing formulas, model error and hedging derivative
portfolios}
\author{T. R. Hurd\thanks{Research supported by the Natural Sciences
and Engineering Research Council of Canada and  MITACS,
Canada}\\Dept. of Mathematics and Statistics\\McMaster
University\\Hamilton, Ontario\\L8S 4K1}
\date{\today}
\maketitle
\begin{abstract}
We propose a method for extending a given asset pricing formula to
account for two additional sources of risk: the risk associated
with future changes in market--calibrated parameters and the remaining
risk associated with idiosyncratic variations in the individual assets
described by the formula. The paper makes simple and natural assumptions
for how these risks behave. These extra risks should always be included
when using the formula as a basis for portfolio management. We
investigate an idealized typical portfolio problem, and argue that a
rational and workable trading strategy can be based on minimizing the
quadratic risk over the time intervals between trades. The example of the
variance gamma pricing formula for equity derivatives is explored, and
the method is seen to yield tractable decision strategies in this case.
\end{abstract}


\section{Introduction}

Any rational and practical approach to managing a financial
portfolio naturally involves finding an optimal balance between a
number of competing criteria.  For example, a manager will wish
to make good (but not excessive) use of available  market data. They
will also need to work with a pricing model in order to estimate
risk  and project to the future. However, the pricing model must
balance complexity and accuracy against simplicity and
implementability. They will want to trade but not too
frequently, and before each trade will need to choose from many
alternatives. They may need a set of investment decision strategies
which are explicable to  their boss, and will stand up to stringent
scrutiny. They must be aware that strategies and parameters alike need to
be updated continuously.

The  paper focuses on an idealized market of derivatives on a single
underlying asset
$S_t$ over a time period $t\in[0,T]$. We address the hypothetical
situation of a manager who has  written (sold) to a client a large and
risky over-the-counter (nontradable) contingent claim $F$ on
$S$, with maturity date $T$ (which we think of as quite large, say
three years). Having accepted the large fee, and the large
liability, the manager must now  create a portfolio of investments
in the exchange traded derivatives with the aim of  adequately or
even optimally hedging the risk. Thus the market consists of a number of
different securities: a bank account, a non--dividend--paying stock, a
family of derivatives on the stock and finally the over--the--counter
claim. We assume
\begin{enumerate}
\item the bank account pays and charges a constant rate of interest $r$,
which we take to be zero for simplicity. There is no
limit on borrowing;
\item the stock price at time $t$, denoted $S_t$, is always
positive;
      \item the exchange traded derivatives are taken to be european
puts and calls over a  variety of strikes $K^\alpha$ and maturities
$T^\alpha$ whose prices at time $t$ are  denoted $D^\alpha_t,
\alpha=2,\dots,M$. By convention, we denote the stock itself
by
$D^\alpha$ with $\alpha=1$.
\item the over--the--counter claim price
$f_t$ is the value at time $t$ of a claim given by a function $F$ of
the stock price history $\{S_t\}_{t\le T}$. That is, $F_T$ is an
$\cF_T$ random variable. By convention, we denote the
over--the--counter claim value by
$D^\alpha$ with $\alpha=0$.
\item the  market is frictionless: there is perfect liquidity, no
transaction costs, unlimited shortselling, and the market is open at all
times.
\end{enumerate}

We assume the manager will set up
an initial portfolio at $t=0$, and then be making single trades at
certain times
$t_1<t_2<\dots<t_k<\dots<T$. In principle, these will be stopping
times, but for this paper we take the deterministic values
$t_i=i\delta t$ for $\delta t$  fixed. Our problem is to provide
this manager  with a consistent, workable strategy for trading, and
to measure the performance of this strategy. The following two
questions are addressed in this paper:
\begin{enumerate}
\item[Q1] What is the condition for optimal allocation at time
$t=0$?
\item[Q2] What is the optimal single (or double or triple) trade which can
be made at time
$t_i, i>0$?
\end{enumerate}

It is supposed that the rational manager will base their strategy on a
derivative pricing model whose historical performance has been studied,
benchmarked, and found to lead to acceptable modeling errors.
This model is taken to be a finite parameter ``risk neutral'' pricing
formula, whose parameters $\theta^a, a=1,\dots,N$ are to be
calibrated to the observed prices of the $M-1$ derivatives (where $N<<M$).
Of course, in contradiction to the modeling assumptions, the calibrated
values
$\hat\theta^a_t$ will depend on the time $t$, reflecting changing market
conditions. Furthermore, the calibrated model will generally fail to
match all observed prices at a given time, the differences being thought
of as ``idiosyncratic errors''. The key point of the paper is to extend
the original model by including minimal additional assumptions on changing
parameters and idiosyncratic errors which are consistent with the observed
performance of the model. These extra effects must be included in any
rational portfolio strategy.

We argue that a rational strategy for the manager will be to
minimize the total {\it risk--neutral} variance of the portfolio returns
over each inter--trading time interval, without regard to the mean
portfolio return. ``Risk--neutral'' refers to the measure which correctly
prices the  current market but not historical values. Variance
optimization is equivalent to optimization with a quadratic utility
function, which is an acceptable approximation to the general utility
function over short time--intervals. Disregarding the mean return is
consistent with a no--arbitrage condition true in an efficient market,
but also justifiable pragmatically because the manager is profiting from
the large fee collected for underwriting the claim
$F$, and needs only consolidate that fee with minimal risk. Furthermore,
estimating mean returns over short time intervals is a game for
speculators and arbitrageurs, not hedgers.

Before each trade the manager will perform a number of steps:
\begin{enumerate}
\item place the payoff of newly expired options (if any) in the bank
account;
\item observe the market prices;
\item  recalibrate the risk--neutral parameters to the new data;
\item estimate the value of the liability $F$ using the recalibrated
pricing model;
\item find the optimal single trade which will minimize the
estimated {\it risk--neutral} variance of the portfolio over the
period until the next trading time.
\end{enumerate}

The main theorem of this paper states that the risk-minimal portfolio
for any trading time-interval is unique, as is the optimal single trade
at any time.

We then discuss models such as the VG model \cite{MaCaCh99} for which the
stock process is Markovian and has a closed form characteristic function.
It is shown that the Fourier transform pricing method of \cite{CarMad00}
extends and allows for efficient computation of portfolio variances, and
hence optimal portfolios and trades.

One important outcome of the present discussion is a clear
operational definition of essential concepts of hedging which shows how
these concepts can be refined and extended: delta hedging, gamma hedging,
vega hedging, model recalibration. Furthermore, the inclusion of
risk associated to ``idiosyncratic errors'' breaks the degeneracy
(nonuniqueness) of the hedging decisions derived from naive
delta--hedging (in practise by biasing trading toward
near--the--money instruments). For a
detailed discussion of hedging and derivatives from the trading
perspective, see
\cite{Taleb97}. For a discussion of model risk, see \cite{Rebon01}. For
another approach to optimal investment in derivatives see \cite{CaJiMa01}.

\section{The model}
\subsection{The option pricing formula} The option pricing formula
is assumed to arise via arbitrage pricing theory set in  the
risk--neutral filtered probability space
$(\Omega,
\cF,\{\cF_t\}_{t\in[0,T]},
\bQ)$ over a time horizon $[0,T]$.  For each
fixed set of parameters
$\theta=(\theta^a)^N_{a=1}\in\bR^N$, the stock is a positive
$\bQ$--martingale (``c\'{a}dl\'{a}g'') (recall the interest rate is taken
to be zero). We consider models in which $S_t$ is one component of a
multidimensional Markov process $(S_t,Y_t)$ whose values are observable
in the market: in our examples $Y$ will have dimensions zero and one. Let
${\cal Y}_t=\sigma\{S_\tau,Y_\tau,\tau\le t\}\subset\cF_t$ be the
filtration of market observables.

The pricing formula giving the value at time $t$  for any
European style contingent claim with payoff
$F_T$ at date $T$ is simply
\be\label{formula}
F(t,S,Y,\theta)=E(F_T(S_T)|{\cal Y}_t,S_t=S,Y_t=Y,\theta\mbox{
fixed}),\quad t\in[0,T]\ee

\nind{\bf Examples: \ }\begin{enumerate}
\item Black--Scholes model: The single
parameter can be taken as $\theta=\log\sigma$ and the stock
process is given by $S_t=S_0\exp[e^\theta
W_t-e^{2\theta}t/2]$ (there are no extra $Y$ variables). The pricing
formula if
$F_T=(S_T-K)^+$ is simply the Black--Scholes call option formula.

\item Variance--Gamma model (\cite{MaCaCh99}) Here the log return process
$X_t=\log\left(S_t/S_0\right)$ is defined to be the L\'evy process
\cite{MR98e:60117},\cite{MR90m:60069}:
\be X_t=\mu_\theta t+\int^t_0\int^\infty_{-\infty}\ x
N^{(\nu_\theta)}(dx dt)\ee   whose jump intensity measure is
\be \nu_\theta(x)=\frac{\alpha e^{-|x|/\eta_\pm}}{|x|},\quad\pm x>0
\ee  for three positive parameters $\theta=(\alpha,\eta_+,\eta_-)$. The
characteristic function turns out to be
\begin{eqnarray}
\Phi_{X_t}(u)&=&E(e^{iuX_t})=\left[\Phi(u)\right]^t\nn\\ \nn\\
\Phi(u)&=&\left[\frac1{(1-i\eta_+ u)(1+i\eta_- u)}\right]^\alpha
\cdot e^{i\mu_\theta u}\label{VGchar}
\end{eqnarray} The martingale condition on $S$ needs $\eta_+<1$ and
is then equivalent to
$\Phi(-i)=1$, so
\be
e^{-\mu_\theta}=\left[\frac1{(1-\eta_+ )(1+\eta_-
)}\right]^\alpha
\ee
    With this condition satisfied, then the stock
process $S_t$ has the form
\be
S_t=S_0+\int^t_0 \int^\infty_{-\infty} (e^x-1) S_{\tau^-} \tilde
N^{(\nu_\theta)}(dx\ d\tau)\ee  where $\tilde N$ denotes the
compensated (martingale) process
\be \tilde N^{(\nu_\theta)}(dx\ dt)= N^{(\nu_\theta)}(dx\
dt)-\nu_\theta(x)\ dx\ dt
\ee

\item Stochastic volatility models:
Standard stochastic volatility models
\cite{Hest93},\cite{PapSir00b}, \cite{PapSir00c},\cite{DuPaSi00}
such as
\begin{eqnarray}
dS_t&=&\sqrt{v_t}\ S_t\ dW^1_t\nn\\
dv_t&=&(a-bv_t)\ dt+\sigma\ \sqrt{v_t}\ dW^2_t\label{stocvol}
\end{eqnarray}
can be considered. This model is Markovian in
$(S_t,v_t)$, not in
$S_t$ alone and  the option pricing formula at time $t$
depends on the values $(S_t,v_t)$. Our approach will be to follow
\cite{BolZho01} and  treat the stochastic squared volatility
$v_t$ as the ``effective observable'' defined by
\be v_t=\lim_{\Delta t\downarrow 0} \lim_{N\to\infty}\ N^{-1}\
\sum_{i=1}^N|\log(S_i/S_{i-1})|^2
\ee  where $S_i\equiv S_{t-(N-i)\Delta t/N}$. This model has four
parameters $(a,b,\sigma,\rho)$ where $d\left<W^1,W^2\right>=\rho dt$.

\end{enumerate}

Many models like the VG model and some stochastic volatility models
have a closed formula for the characteristic function
$\Phi_{X_t}$. In such cases \cite{CarMad00} have shown how the Fast
Fourier transform method provides a numerically efficient method for
evaluating (\ref{formula}). For example, the European call payoff
function with a strike $K=e^{k}$ can be written as a
complex contour integral along the  shifted contour
$(-\infty-i\epsilon,\infty-i\epsilon)$ for any
$\epsilon>0$:
\be\label{fourier}
(S-e^k)^+=\frac{1}{2\pi}\int^{\infty-i\epsilon}_{-\infty-i\epsilon}
\ S^{iu+1}\ e^{-iuk}Q(u)\ du
\ee
where
\be   Q(u)=\frac 1{iu-u^2}
\ee
To take the expectation to evaluate (\ref{formula}), plug in this formula
and interchange the
$du$ and $E(\cdot)$ integrals
\begin{eqnarray}
F(t,S)&=&\frac{1}{2\pi}\int^{\infty-i\epsilon}_{-\infty-i\epsilon}
E\left(S_T^{iu+1}\
e^{i(u-i)X_{T-t}}\ e^{-iuk}Q(u)\right)\ du\nn\\ \nn\\
&=&\frac{1}{2\pi}\int^{\infty-i\epsilon}_{-\infty-i\epsilon}
\ S_t^{iu+1}\
\Phi(u-i)^{T-t}\ e^{-iuk}Q(u)\ du
\end{eqnarray}
The explicit integral can now be evaluated (approximately) very
efficiently for a linearly spaced family of log strike values
$\{k_0+n\delta k\}_{n=0,\pm1,\pm 2,\dots}$ using the fast Fourier
transform. Similar formulas hold for put options and the
over--the--counter claim itself if its payoff has a known Fourier
transform.

\subsection{Varying market conditions} A careful trader using a pricing
formula such as (\ref{formula}) will wish to correct for changing market
conditions. It will be supposed that prior to its practical application
(to hedging) the performance of the formula has been carefully
benchmarked against real market data and a historical time series of mean
square estimated values
$\hat\theta_{t}$ (as described in section
\S 3.1) has been observed.
When we allow the vector of parameters
$\theta$ to become a process, it is natural to pick a gaussian
mean--reverting process and therefore we assume:
\begin{enumerate}
\item[{\bf A1}] The time varying parameters $\theta^i_t$ form an
$\Reals^N$--valued Ornstein--Uhlenbeck Ito diffusion (gaussian, mean
reverting). The noise process driving
$\theta_t$ is independent of the noise process driving $S_t$.
\end{enumerate}
Let $\cG_t=\sigma\{\theta_s,s\le t\}\subset\cF_t$.

A consistent pricing formula based on the filtration
${\cal Y}_t\times\cG_t$ is given by
\be \tilde F (t,S,Y,\theta)= E(
F_T(S_T)|{\cal Y}_t\times\cG_t,S_t=S,Y_t=Y,\theta_t=\theta),\quad
t\in[0,T]\ee Note that $\tilde F $ depends on the detailed modeling of the
process (described of course by extra parameters), and hence
$\tilde F \neq F $.
For the present paper we regard use of $\tilde F$ in place of $F$
as a change in the underlying pricing model, and is hence ``against
the rules''.

Since our purpose is rather to use $ F $ to find a hedging strategy which
will include low order approximations to account for
variations in
$\theta$, the point of view of varying parameters can be
simplified. We adopt a natural compromise position which is to regard
the parameters $\theta$ as constant over the intertrading intervals
$[t_{i-1},t_{i})$. At each time $t_i$ the parameters are assumed to
jump according to a {\it discrete} $\Reals^N$-valued OU process which is
independent of the stock price $S_{t_i}$.  Over the
time period $[t_{i-1},t_{i})$, the stock price is assumed to evolve
according to the model with fixed
$\theta=\theta_{t_{i-1}}$. The distribution of $\theta_{t_i}$ conditioned
on
$\theta_{t_{i-1}}$ is gaussian. To keep the discussion simple, we will
focus on the conditional covariance estimated by the quadratic
variation statistic:
\be
\widehat
\Theta^{ab}=\frac1{n}\sum_{k=1}^n[\hat\theta^a_{\tilde
t_k}-\hat\theta_{\tilde t_{k-1}}^a] [\hat\theta^b_{\tilde
t_k}-\hat\theta_{\tilde t_{k-1}}^b],\quad a,b=1,\dots N
\ee
obtained from a series of times
$\tilde t_0,\tilde t_1,\dots,\tilde t_n$ (all prior to the trading times)
with
$\tilde t_i-\tilde t_{i-1}=\delta t$.

\subsection{Idiosyncratic variations} Even if the model parameters
$\theta$ are confidently known at time
$t$, the observed option prices $D^\alpha$ will not fit the formula
perfectly. There are certainly model errors which are idiosyncratic
to each exchange traded security, and a careful trader must also
account for this. As before, it is supposed that these
idiosyncratic errors have been benchmarked by observing how
the formula matches real data and that the process
\be\label{idio} \cE^\alpha_t= D^\alpha_{\tilde t} - F
^\alpha( t,S_{ t},\theta_t)
\ee
has been sampled over time.
We assume \begin{enumerate}
\item[\bf{A2}] the errors $\cE_t$  form a multidimensional
Ornstein--Uhlenbeck Ito diffusion, independent of the noise processes
driving
$S_t$ and $\theta_t$. For simplicity, we assume that the covariance of
$\cE_{t_i}$ conditioned on $\cE_{t_{i-1}}$ is diagonal.
\end{enumerate}

Trading
instincts suggest that the predicted variances will depend most strongly
on two variables, namely moneyness
$\kappa=K/S$ and time to maturity
$\tau=T-t$, both of which vary in time for a specific derivative
$\alpha$. Therefore we adopt the following crude method for predicting the
idiosyncratic variances:  Let the variance of $\cE^\alpha_{t+\delta t}$
conditioned on
$\cE^\alpha_{t}$ be
\be \hat V^\alpha_t=S_t^2\hat v(K^\alpha/S_t,T^\alpha-t)\ee  where $\hat v$
is the following historically observed quantity which depends on moneyness
     and time to maturity:
\be \hat
v(\kappa,\tau)=\frac1{n\delta t}\sum_{k=1}^n S_{\tilde t_{k}}^{-2}
[\hat\cE^{\alpha(\tilde t_k,\kappa,\tau)}_{\tilde t_k+\delta
t}-\hat\cE^{\alpha(\tilde t_k,\kappa,\tau)}_{\tilde t_k}]^2
\ee
where $\{\tilde t_k\}$ is a sequence of times prior to the trading times
$t_i$. Here
$\alpha(t,\kappa,\tau)$ is defined to be that derivative whose
characteristics
$(K^\alpha,T^\alpha)$ at time
$t$ most nearly match
$(\kappa,\tau)$, i.e. the
$\alpha$ which minimizes
\be
d(\alpha;(\kappa,\tau))=(K^\alpha/S_t-\kappa)^2+((T^\alpha-t)/\tau-1)^2\ee

\section{Optimal trading} We now develop in detail the
problem of optimal trading at time $t_i$. In what follows we use the
shorthand notation
$S_i\equiv S_{t_i}$, etc. for quantities evaluated at
time
$t_i$.
The formulation we will adopt will take into account the
following information at time $t_i$: the observed market prices
$D^\alpha_i$, the empirical covariance $\widehat \Theta^{ab}$, and the
empirical variances $\hat V^\alpha_i$.

We assume that immediately following the previous trading time
$t_{i-1}$  the (self--financing) portfolio consists of
$\pi^\alpha_{i-1}$ units of each derivative, the liability for the
claim
$D^0$ (so
$\pi^0=-1$ always), and the remaining value $B_{i-1}$ in the
bank account. The portfolio value $X$ is
\begin{eqnarray} X_{i-1}&=&\sum^M_{\alpha=0} \pi^\alpha_{i-1}
D^\alpha_{i-1}  + B_{i-1}\\ B_{i-1}&=& X_{{i-1}}-\sum_\alpha
\pi^\alpha_{i-1}D^\alpha_{{i-1}}
\end{eqnarray}

\subsection{Calibration to observed prices} At time $t_i$, we
compare the observed market prices $D^\alpha_i, \alpha=1,\dots,N$  to the
model prices to update the market parameters $\theta$. Following
\cite{MaCaCh99}, the vector of estimated parameters
$\hat\theta_{i}$ is the minimizer of
\be\label{opt}
  \min_{\theta}\{\sum_\alpha\  |\log( D^\alpha_i)-
     \log( F ^\alpha(t_i,S_i,\theta))|^2\}\ee  Note that by definition
$D^0= F ^0$ and $D^1= F ^1=S$ so the sum is over $\alpha=2,\dots,
N$. Refining (\ref{opt}) by weighting by trading volume might be
considered.

\subsection{Marking to market} Given the
recalibrated market parameters
$\hat\theta_i$ and current market prices $D^\alpha_i,\alpha\ge 1$,
the updated portfolio value can be calculated for time $t_i$. The best
estimate of the current value of the contingent claim $D^0$ is the
expected value at $t_i$ of the (discounted) final claim:
\be D^0_i= F ^0(t_i,S_i,\hat\theta_i)\ee  (this explains
why $\cE^0=0$ always).
Therefore the updated portfolio value, just prior to trading, can be
written\be  X_i=\sum_{\alpha \ge 0}\pi^\alpha_{i-1}
      F ^\alpha(t_i,S_i,\hat\theta_i) +\sum_{\alpha\ge 2}
\pi^\alpha_{i-1}  \cE_i^\alpha + B_{i-1}\ee

\subsection{Finding the optimal trade} To determine the optimal
trade (i.e. to find the best new portfolio weights
$\pi^\alpha_{i}, B_{i}$), we project
\be  X_i=\sum_{\alpha \ge 0}\pi^\alpha_{i}
      F ^\alpha(t_i,S_i,\hat\theta_i) +\sum_{\alpha\ge 2}
\pi^\alpha_{i}  \cE_i^\alpha + B_{i}\ee forward to the next trading
time
$t_{i+1}$ and consider the (stochastic) value
$X_{{i+1}}$. This is characterized by the values $\pi^\alpha_{i},
B_{i}$, the new market prices $D^\alpha_{{i+1}}
$, and the new parameter values $\hat\theta_{{i+1}}$. Consider the
difference $\De X\equiv X_{{i+1}}-X_i=\sum_{\alpha \ge 0}\pi^\alpha_{i}
\Delta X^\alpha$ where
\begin{eqnarray*}
\Delta X^\alpha&=&
[ F^\alpha(t_{i+1},S_{{i+1}},\hat\theta_{i})-
    F^\alpha(t_{i},S_{{i}},\hat\theta_i)
]\\
&&+[ F^\alpha(t_{i+1},S_{{i+1}},\hat\theta_{i+1})-
    F^\alpha(t_{i+1},S_{{i+1}},\hat\theta_i)
] +[\cE^\alpha_{{i+1}}-\cE^\alpha_{i}]\\&\equiv&
\Delta X_1^\alpha+\Delta X_2^\alpha+\Delta X_3^\alpha\end{eqnarray*}

The middle term $\Delta X_2$ will become a bit awkward, so we reorganize
it by Taylor expanding in powers of
$\widehat {\Delta
\theta_i}=\hat\theta_{i+1}-\hat\theta_i$:
\be
F^\alpha(t_{i+1},S_{i+1},\hat\theta_{i+1})-
F^\alpha(t_{i+1},S_{i+1},\hat\theta_i)
=\sum_{n\ge
1}(\partial_\theta^n F^\alpha)(t_{i+1},S_{i+1},\hat\theta_{i})
\widehat {\Delta
\theta_i}^n
\ee
We have adopted a condensed multiindex notation so that
$\partial_\theta^n\equiv \pa_{\theta^1}^{n_1}\dots\pa_{\theta^N}^{n_N}$
and $\widehat {\Delta
\theta}^n\equiv
{(\hat\theta^1)}^{n_1}\dots{(\hat\theta^N)}^{n_N}$ with
$n=(n_1,\dots,n_N)$. $n\ge 1$ is shorthand for $\sum_i n_i\ge 1$.

    In what follows, let $E_i(\cdot)$ denote conditional
expectations
$E_Q(\cdot|\cF_{t_i})$. We consider first the mean of $\Delta X$.
Note $E_i(\De X^\alpha_1)=0$ since $F^\alpha$ is a martingale. In
general, however,
$E_i(\De X^\alpha_2)\neq 0, E_i(\De X^\alpha_3)\ne 0$ since conditional
means of OU processes are never identically  zero. Statistically, mean
returns are small, difficult to measure and very unstable in
time.  From a trading point--of-view, seeking mean
returns over short time intervals is the job of the speculator, the
day--trader and the arbitrageur.  For our hedger, it makes no sense  to
trade on expected returns.
Therefore, from both the statistical and trading points of view it makes
sense to introduce an extra assumption
\begin{enumerate}
\item[{\bf A3}] The trader will ignore the mean returns of
$\Delta X_2, \Delta X_3$, and will trade as if
$E_i(\De X^\alpha_2)= 0$ and $E_i(\De X^\alpha_3)= 0$.
\end{enumerate}

Thus we assume that when conditioned on $\cF_i$,  $\widehat
{\Delta \theta_i}$ and $\cE^\alpha_{{i+1}}-\cE^\alpha_{i}$ are taken to be
mean zero gaussians, with conditional covariances $\widehat\Theta^{ab}$
and
  $\hat V^\alpha_i$.

Under these assumptions,  the variance or total quadratic
risk $$R\equiv\var_i(\Delta X)=\sum_{\alpha\beta}\pi^\alpha_i\pi^\beta_i
R^{\alpha\beta}$$ decomposes into three terms
\begin{eqnarray*}
R^{\alpha\beta}&\equiv& {\bf
\mbox{ Cov}}_i(\Delta X_1^\alpha;\Delta X_1^\beta)+{\bf\mbox{
Cov}}_i(\Delta X_2^\alpha;\Delta X_2^\beta)\\&&+ {\bf\mbox{
Cov}}_i(\Delta X_3^\alpha;\Delta X_3^\beta)\\\\
&=&R_1^{\alpha\beta}+R_2^{\alpha\beta}+R_3^{\alpha\beta}
\end{eqnarray*}
\begin{itemize}
\item $R_1$ is the risk associated with changes in the underlier $S$;
\item $R_2$ is the risk associated with changes in the market
reflected in evolving parameter values;
\item $R_3$ is the risk associated with deviations
between observed prices and model prices.
\end{itemize}
This type of decomposition is
quite general.

We have now succeeded in expressing the quadratic portfolio risk in terms
of the pricing formulas $ F ^\alpha$, the conditional covariance
matrix $\widehat\Theta^{ab}$ and the variances $\hat V^\alpha$.

\bpr Consider a pricing model of the above type under assumptions
{\bf A1,A2,A3}. Then
\begin{enumerate}
\item there is a   unique portfolio $\pi_i^*$ with $\pi_i^0=-1$ which
minimizes the quadratic risk
\be
\pi_i^*={\arg\min}_{\pi:\pi^0=-1}\sum_{\alpha,\beta\ge
0}\pi^\alpha\pi^\beta [R^{\alpha\beta}_1+ R^{\alpha\beta}_2+
R^{\alpha\beta}_3]\ee
over the period $[t_i,t_{i+1}]$ given by the solution of
\be
R^{0\alpha}=\sum_{\beta>0}R^{\alpha\beta}\pi_i^{*\beta},\alpha>0
\ee
\item for any portfolio $\pi_{i-1}$ with $ \pi^0=-1$, there is a unique
optimal single trade consisting of $\lambda^*$ units of the
$\gamma^*$th derivative, $\gamma^*\ge 1$ where:
\be (\lambda^*,\gamma^*)={\arg\min}_{\lambda,\gamma\ge 1}
\sum_{\alpha,\beta\ge 0}(\pi^\alpha_{i-1}+\lambda\delta^\alpha_\gamma)
(\pi^\beta_{i-1}+\lambda\delta^\beta_\gamma)
[R^{\alpha\beta}_1+ R^{\alpha\beta}_2+
R^{\alpha\beta}_3]
\ee
\end{enumerate}
Here $\arg\min$ denotes the solution of the associated minimization
problem. \epr

\Proof We need only note that $R_1$ is positive definite while
$R_2,R_3$ are positive semidefinite matrices.

\QED

Selecting the optimal single trade given $\pi_{i-1}$ is a simple matter
given the matrix $R$. The optimal amount $\lambda^{*,\gamma}$ to trade of
a given derivative $\gamma$ is
\be
\lambda^{*,\gamma}=-\frac{\sum_{\beta}
R^{\gamma\beta}\pi_{i-1}^\beta}{R^{\gamma\gamma}}
\ee
Such a trade decreases the unimproved portfolio risk
$\sum_{\alpha,\beta}\pi^\alpha_{i-1}\pi^\beta_{i-1}
R^{\alpha\beta}$
by the amount $(\sum_{\beta}
R^{\gamma\beta}\pi_{i-1}^\beta)^2/{R^{\gamma\gamma}}$; the optimal asset
to trade is simply that $\gamma^*$ which shows the maximal improvement.

Similar arguments give one the optimal double, triple trade etc.

\section{Implementing the method}

Implementation of the method for a particular trading formula depends on
efficient determination of the risk matrix $R^{\alpha\beta}$ before
every trade. We demonstrate how this can be done by generalizing the Fast
Fourier Transform (FFT) method of \cite{CarMad00}  when the log-return
process
$X_t=\log(S_t/S_0)$ has a known characteristic function $\Phi(u)$. We
suppose that for each $\alpha=0,1,\dots, M$ the payoff function
$F^\alpha(S_T)$ has a Fourier formula similar to  (\ref{fourier}) but
with the function  $Q(u)$ replaced by $Q^\alpha(u)$.

First note that $R_3$ is always simply the diagonal matrix
\be R_3^{\alpha\beta}=\delta^{\alpha\beta} \hat V^\alpha_i
=\delta^{\alpha\beta}  S_i^2\hat v(K^\alpha/S_i,T^\alpha-t_i)\ee

$R_1,R_2$ are more complicated. We note that they are symmetric matrices
which fall naturally into block submatrices
\be
R=\left[\begin{array}{cccc}
R_{00}&R_{01}&R_{0c}&R_{0p}\\
R_{10}&R_{11}&R_{1c}&R_{1p}\\
R_{c0}&R_{c1}&R_{cc}&R_{cp}\\
R_{p0}&R_{p1}&R_{pc}&R_{pp}
\end{array} \right]
\ee
where $R_{0c}$ is $1\times M_c$,$R_{cp}$ is $M_c\times M_p$ etc. where
$M_c,M_p$ are the number of traded calls and puts. Put--call parity
leads to a number of relations amongst the components:
\be
R_{cc}-R_{cp}=R_{c1},\quad
R_{cp}-R_{pp}=R_{1p},\quad\mbox{etc.}\ee
We
indicate here how to compute the parts of
$R_{cc}$ with fixed maturities
$T^\alpha,T^\beta$ and any log strikes $k^\alpha,k^\beta$.

For $R_1$ terms note that
\begin{eqnarray}
R_{1,cc}^{\alpha\beta} &=& E_i\left([F^\alpha_{i+1}-F^\alpha_i]
[F^\beta_{i+1}-F^\beta_i]\right)\nn \\
&=&
E_i\left(F^\alpha_{{i+1}}F^\beta_{i+1}\right)-
F^\alpha_i
F^\beta_i
\end{eqnarray}
The usual approach is to Taylor expand in powers of $\Delta t$ which
leads to generalized delta--gamma hedging. However with the FFT method, we
can perform the one remaining expectation leading to an explicit double
Fourier integral
\begin{eqnarray}
R_{1,cc}^{\alpha\beta} &=&
\frac1{(2\pi)^2}\int\int
S_i^{i(u_1+u_2)+2}
e^{-iu_1k^\alpha-iu_2k^\beta}Q(u_1)Q(u_2)\nn \\ &&\times
\left[\Phi(u_1+u_2-2i)^{\Delta
t}-\Phi(u_1-i)^{\Delta
t}\Phi(u_2-i)^{\Delta
t}\right]\nn
\\&&\times\Phi(u_1-i)^{T^\alpha-t_{i+1}}\
\Phi(u_2-i)^{T^\beta-t_{i+1}}\ du_1\ du_2\label{R1int}
\end{eqnarray}
A single application of the two dimensional FFT solves this problem for a
family of calls with linearly spaced log strikes but with fixed dates
$T^\alpha,T^\beta$.

Now we seek a similar formula for
\begin{eqnarray}
    R_{2,cc}^{\alpha\beta} &=&
E_i\left[\left(F^\alpha_{i+1}(\hat\theta_{i+1})-
F^\alpha_{i+1}(\hat\theta_{i})\right)
\left(F^\beta_{i+1}(\hat\theta_{i+1})-F^\beta_{i+1}(\hat\theta_{i})
\right)\right]\nn\\
&&-E_i\left[F^\alpha_{i+1}(\hat\theta_{i+1})-F^\alpha_{i+1}(\hat\theta_{i})\right]
E_i\left[
F^\beta_{i+1}(\hat\theta_{i+1})-F^\beta_{i+1}(\hat\theta_{i})\right]
\end{eqnarray}
When we plug in the Fourier integral, we highlight the
$\hat\theta$--dependence of $\Phi$ by writing $\Phi(u;\hat\theta)$. Then
using the independence of $S$ and $\theta$ expectations leads to
\begin{eqnarray}
R_{2,cc}^{\alpha\beta} &=&
\frac1{(2\pi)^2}\int\int
e^{-iu_1k^\alpha-iu_2k^\beta}Q(u_1)Q(u_2)\nn \\ &&
\times
E_i\Biggl(S_{i+1}^{i(u_1+u_2)+2}\
\left[\Phi(u_1-i;\hat\theta_{i+1})^{T^\alpha-t_{i+1}}-
\Phi(u_1-i;\hat\theta_{i})^{T^\alpha-t_{i+1}}\right]\nn\\
&&\times
\left[\Phi(u_2-i;\hat\theta_{i+1})^{T^\beta-t_{i+1}}-
\Phi(u_2-i;\hat\theta_{i})^{T^\beta-t_{i+1}}\right]\Biggr) du_1\
du_2\nn\\ \nn\\ &-&\frac1{(2\pi)^2}\int\int
e^{-iu_1k^\alpha-iu_2k^\beta}Q(u_1)Q(u_2)\nn \\ &&
\times
E_i\Biggl(S_{i+1}^{iu_1+1}\
\left[\Phi(u_1-i;\hat\theta_{i+1})^{T^\alpha-t_{i+1}}-
\Phi(u_1-i;\hat\theta_{i})^{T^\alpha-t_{i+1}}\right]\Biggr)\nn\\
&&\times
E_i\Biggl(S_{i+1}^{iu_2+1}\
\left[\Phi(u_2-i;\hat\theta_{i+1})^{T^\beta-t_{i+1}}-
\Phi(u_2-i;\hat\theta_{i})^{T^\beta-t_{i+1}}\right]\Biggr)\ du_1\
du_2
\end{eqnarray}
which becomes
\begin{eqnarray}
&=&
\frac1{(2\pi)^2}\int\int
S_i^{i(u_1+u_2)+2}
e^{-iu_1k^\alpha-iu_2k^\beta}Q(u_1)Q(u_2)\ \nn\\
&&\Biggl(\Phi(u_1+u_2-2i;\hat\theta_i)^{\Delta
t}
E_i\Bigl[\left[\Phi(u_1-i;\hat\theta_{i+1})^{T^\alpha-t_{i+1}}-
\Phi(u_1-i;\hat\theta_{i})^{T^\alpha-t_{i+1}}\right]\nn\\
&&\times
\left[\Phi(u_2-i;\hat\theta_{i+1})^{T^\beta-t_{i+1}}-
\Phi(u_2-i;\hat\theta_{i})^{T^\beta-t_{i+1}}\right]\Bigr]
\nn\\
&-&\Phi(u_1-i;\hat\theta_i)^{\Delta
t} \Phi(u_2-i;\hat\theta_i)^{\Delta
t}
E_i\left[\Phi(u_1-i;\hat\theta_{i+1})^{T^\alpha-t_{i+1}}-
\Phi(u_1-i;\hat\theta_{i})^{T^\alpha-t_{i+1}}\right]\nn\\
&&\times
E_i\left[\Phi(u_2-i;\hat\theta_{i+1})^{T^\beta-t_{i+1}}-
\Phi(u_2-i;\hat\theta_{i})^{T^\beta-t_{i+1}}\right]\Biggr)\ du_1\
du_2\label{R2int}
\end{eqnarray}
The remaining $\theta$--expectations will be quite complicated since they
involve differencing the function $\Phi$. A pragmatic approach is simply
to Taylor expand in powers of $\widehat{\Delta\theta}$. Then the
leading term is
\begin{eqnarray}
&&\hspace{-.4in}
\frac1{(2\pi)^2}\int\int
S_i^{i(u_1+u_2)+2}
e^{-iu_1k^\alpha-iu_2k^\beta}Q(u_1)Q(u_2)\ 
\Phi(u_1+u_2-2i;\hat\theta_i)^{\Delta
t} \times\nn\\
&&\hspace{.5in}
\sum_{a,b=1}^N\left[
\pa_{\theta^a}(\Phi(u_1-i;\hat\theta_{i})^{T^\alpha-t_{i+1}})
\pa_{\theta^b}(\Phi(u_2-i;\hat\theta_{i})^{T^\beta-t_{i+1}})
\widehat\Theta^{ab}\right]
du_1\ du_2\label{R22int}
\end{eqnarray}
with higher order terms given by more complicated integrals. These
integrals are efficiently calculated by FFT as before.

\subsection{Example: the Black--Scholes model}

The Fourier technique above applies to this model, since the
characteristic function is simply
\be
\Phi(u)=\exp[{-e^{2\theta}(u^2+iu)/2}]
\ee

The matrices $R_1,R_2,R_3$ can be interpreted in the standard language of
hedging. For example, by taking the second order Taylor approximation to
$R_1$ in powers of
$\Delta t$ and setting
$\sum_{\alpha,\beta}R_1^{\alpha\beta}\pi^\alpha\pi^\beta=0$, we obtain
the conditions for a delta--gamma--hedged portfolio. Note that in
our present formulation, the optimal solutions make this term small but
not zero. Similarly, setting the lowest order $\Delta t$ term of $R_2$
equal to zero yields a vega--hedged portfolio which shows vanishing
first--order sensitivity to changes in the estimated volatility. Again
our optimal portfolios will be approximately but not  perfectly
vega--hedged. Finally, making
$R_3$ small amounts to choosing preferentially amongst the more liquid
derivatives, since their idiosyncratic risks are generally relatively
smallest. In practise, this means favoring investment in near--the--money
assets to far in/out--of--the--money derivatives.

We illustrate the method with an example which shows the
optimal hedge for
  a Euro--call contract with strike $1$
and maturity
$T=3$ (all times in years) whose total Black--Scholes value is $\$1$. In a
pure Black--Scholes market where the current stock price is
$\$1$, with puts and calls with maturity
$1/2$ and 11 log strikes
$k=(-0.5,-0.6,\dots,1.5)$, we calculated the optimal hedge
(excluding the underlier $S$) for the period $\de t=1/12$. For
illustrative purposes, we take the other parameters to be $
\widehat\Theta=10^{-2},\hat v_c=\hat v_p=10^{-3}(6,5,4,3,2,1,2,3,4,5,6)$.
The graph shows the value bought of each call option ($+$) and the value
($\times$) sold of each put. The ratio of quadratic hedged risk to
unhedged risk is $0.0319$, the hedged to unhedged delta ratio  is
$-0.0441$ and the hedged to unhedged gamma ratio is $-0.7411$. With
optimized choices for FFT parameters the computation required  $21.3$
megaflops.

\centerline{\epsfxsize=5in\epsfbox{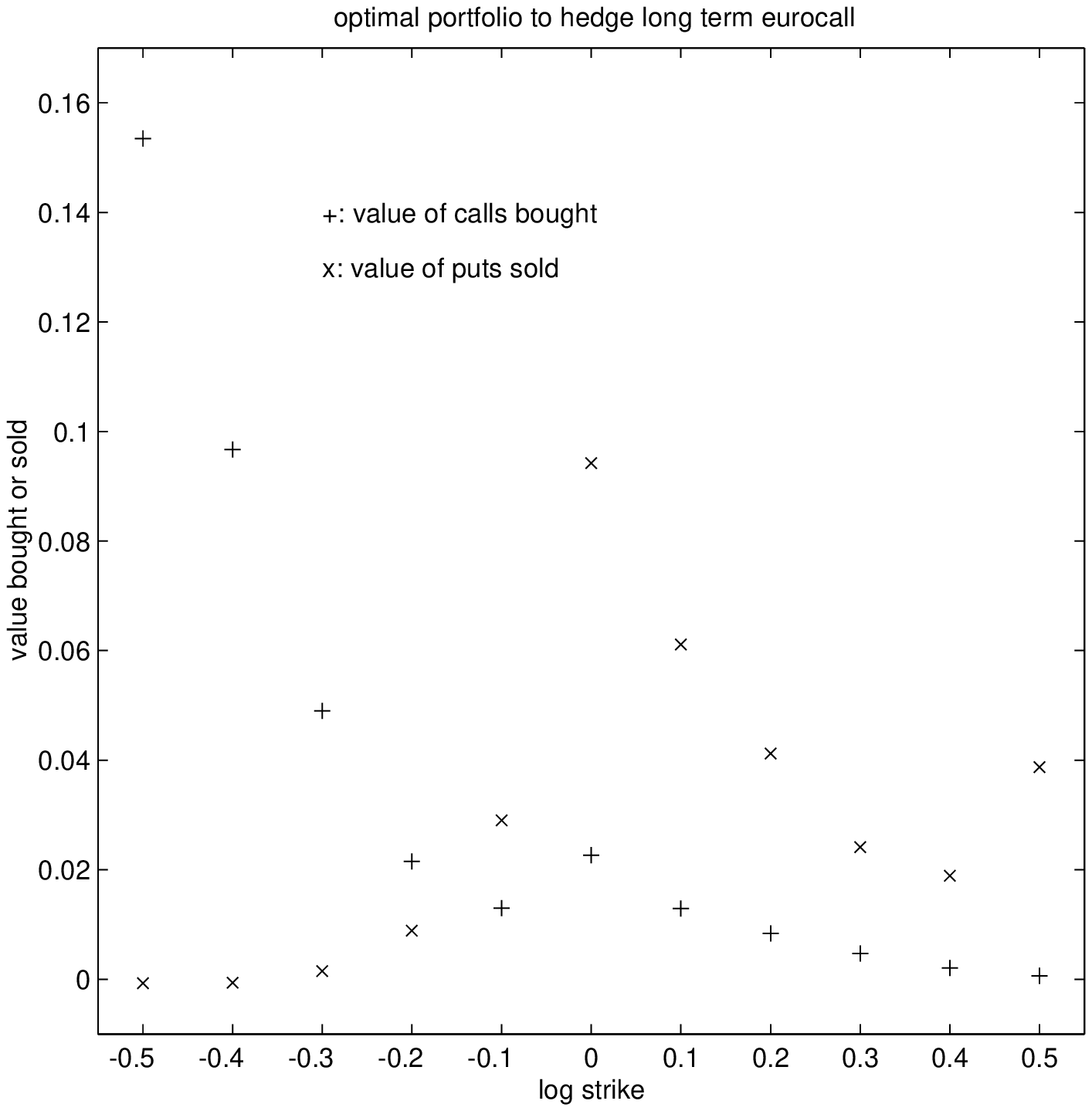}}

\subsection{Example: the VG model}

We begin by noting that for this model the parameterization
$\hat\theta=(\alpha,\eta_+,\eta_-)$ is not a linear space, but can be
associated with a natural riemannian manifold with metric given by the
Fisher information metric. Therefore, we replace assumption {\bf
A1} by the natural family of Ornstein-Uhlenbeck processes
which constitute a preferred class of mean--reverting processes on a
general riemannian manifold. However,
we expect this nontrivial geometry to affect only the higher order
$\widehat{\Delta\theta}$ terms in the Taylor expansion for $R_2$ and for
a preliminary implementation we would work in the tangent space and keep
only first order Taylor terms. Thus we would use (\ref{R22int}) with the
estimated
$\widehat\Theta$ matrix as before.

Now, since the characteristic function for the VG model has the
simple analytic expression (\ref{VGchar}), the important double integrals
can be efficiently calculated via the FFT.

It is now known that the VG model does not work well in pricing
derivatives  across maturities, and so our framework cannot be expected
to perform adequately. Work is in progress to extend the VG model to
include nontrivial time correlations.

\subsection{Example: the  affine stochastic volatility
model}

In the model given by equation (\ref{stocvol})  the
characteristic function $\Phi(t,u)=E(e^{iX_tu})$ for
$X_t=\log(S_t/S_0)$ depends on the current level of squared volatility
$v_0$:
\be
\Phi(t,u)=\exp[C(t,u)+D(t,u)v_0]
\ee
where
\begin{eqnarray}
C(t,u)&=&\frac{a}{\sigma^2}\left[(b-i\rho\sigma
u+d)t-2\log\left(\frac{1-ge^{dt}}{1-g}\right)\right]\nn\\
D(t,u)&=&\frac{b-i\rho\sigma
u+d}{\sigma^2}\left[\frac{1-e^{dt}}{1-ge^{dt}}\right]\nn\\
g&=&\frac{b-i\rho
\sigma u +d}{b-i\rho
\sigma u -d}\nn\\
d&=&\sqrt{(i\rho\sigma u-b)^2-\sigma^2(-iu-u^2)}\
\end{eqnarray}
where $\rho$ is the correlation between $W^1,W^2$. Our general method
still applies, however the dependence on
$v_0$ leads to more complicated integrals for
$R_1,R_2$ which are deserving of further study.

\section{Conclusions}

The general picture is that a finite parameter pricing model can be
augmented by an assumption that parameters vary stochastically
(hopefully slowly) in time, independently of other sources of
randomness. Furthermore, one need not assume idiosyncratic errors
are zero: these too can be modeled by a simple independent random
process. These assumptions can be built naturally into hedging
strategies which minimize risk over a given time interval between
trades. This risk naturally decomposes into a sum of three terms: a
term involving changes in the underlier, a term involving changes
in the parameters and a term involving the idiosyncratic errors.
Expectations involving changes in the underlier are with respect to
the risk--neutral measure, and thus incorporate a non-zero price of
risk, whereas the remaining probabilities are with respect to the
physical (historical) probabilities. Decision making for the
resulting hedging strategy at a given trading time requires knowing
how to calculate model prices and generalized greeks for fixed values of
the parameters (using the analytical formulas, Monte Carlo or some other
method) and knowing the following data:
\begin{enumerate}
\item the current prices of all exchange traded securities;
\item historically estimated unconditional variances for the
changes in parameters and idiosyncratic pricing errors.
\end{enumerate}

The strategy which results has a number of practical advantages:
\begin{itemize}
\item It provides a systematic treatment of model errors, hence leading
to smaller absolute hedging errors;
\item Since decision making depends only on current market prices and
benchmark statistics which are relatively robust and stable in
time, we expect hedging errors over successive trading
intervals to exhibit only small statistical dependence. This means that
the probability distribution of the hedging error over longer terms can be
estimated by the central limit theorem;
\item The method can be implemented efficiently in a wide variety of
models, both simple and sophisticated, which include pure diffusions, jump
diffusions and pure jump models.
\end{itemize}

It should be observed once more that we are introducing new parameters
(here $\hat V, \widehat\Theta$) which describe idiosyncratic errors and
the changes in model parameters. Of course these themselves are best
estimated dynamically in time. Thus the finicky practitioner might be
lead to introduce further parameters which describe the changes in the
parameters
   which describe the changes
in model parameters, etc! The logic of the paper still applies.

Further work is suggested along a number of lines:

\begin{itemize}
\item  A comparison of
the performance of our method relative to the standard
discrete time hedging strategies;
\item The statistical dependence of successive hedging errors should be
studied and compared to that of other strategies;
\item A study of whether the strategy is consistent with observed
trading patterns in the market or whether on the contrary traders
use different criteria in decision making;
\item In the current paper we
avoid discussing transaction costs by assuming a fixed schedule of single
trades. The work should be extended to include transaction costs, and
which will allow questions of when it is optimal to trade, and when
double or higher order trades are preferable over single trades;
\item The method should
also be applied to models for other markets, such as the commodity, bond
and foreign exchange markets.
\end{itemize}

To conclude, the current paper is not intended to present a
mathematically consistent model. On the contrary, it addresses the
question of how any mathematically consistent pricing model can be
extended for use by a rational and prudent practitioner who cares about
the inevitable errors made by the model.

\end{document}